\documentclass[useAMS,usegraphicx,usenatbib]{mn2e}

\title[Beryllium abundances in metal-poor stars]{Beryllium abundances in metal-poor
stars \thanks{Based on observations made with the European Southern Observatory
telescopes obtained from the ESO/ST-ECF Science Archive Facility.}}
\author[K. F. Tan, J. R. Shi and G. Zhao]{K. F. Tan$^{1,2}$\thanks{E-mail: tan@bao.ac.cn},
J. R. Shi$^{1}$ and G. Zhao$^{1}$\\
$^{1}$National Astronomical Observatories, Chinese Academy of Sciences, Beijing 100012, China\\
$^{2}$Graduate University of the Chinese Academy of Sciences, Beijing 100049, China}

\begin{document}

\date{Accepted ?. Received ?; in original form ?}

\pagerange{\pageref{firstpage}--\pageref{lastpage}} \pubyear{2008}

\label{firstpage}

\maketitle

\begin{abstract}
We have determined beryllium abundances for 25 metal-poor stars
based on the high resolution and high signal-to-noise ratio spectra
from the VLT/UVES database. Our results confirm that Be abundances
increase with Fe, supporting the global enrichment of Be in the
Galaxy. Oxygen abundances based on \mbox{[O\,{\sc i}]} forbidden
line implies a linear relation with a slope close to one for the Be
vs. O trend, which indicates that Be is probably produced in a
primary process. Some strong evidences are found for the intrinsic
dispersion of Be abundances at a given metallicity. The
deviation of HD\,132475 and HD\,126681 from the general Be vs. Fe
and Be vs. O trend favours the predictions of the superbubble model,
though the possibility that such dispersion originates from the
inhomogeneous enrichment in Fe and O of the protogalactic gas cannot
be excluded.
\end{abstract}


\section{Introduction}\label{intro}

The light element beryllium (Be) has very special origins. The
primordial Be abundance (on the order of $N\rmn{(Be/H)}=10^{-17}$,
\citealt{thomas1994}) is negligible as predicted by the standard Big
Bang Nucleosynthesis. Be can not be produced by nuclear fusion in
the interiors of stars; on the opposite, Be will be destroyed by
this process. It was first proposed by \citet*{reeves1970} that Be
can be created by spallation reactions between galactic cosmic rays
(GCRs) and the CNO nuclei in the interstellar medium (ISM). This
model predicts a quadratic relation between the abundances of Be
and O (or a slope of 2 in logarithmic plane), assuming that the CNO
abundance is proportional to the cumulative number of Type II
supernovae (SNe\,II) and the cosmic ray flux is proportional to the
instantaneous rate of SNe\,II. However, recent observational results
(e.g., \citealt{gilmore1992, boesgaard1993,molaro1997}) showed a
linear relation between Be and O abundances, which indicates that Be
may be produced in a primary process instead of the standard
secondary GCRs process. \citet{duncan1997} suggested that Be can be
produced in a reverse spallation of C and O nuclei onto protons and
$\alpha$-particles. This process will lead to a linear dependence of
Be on O abundances. The results from the latest big survey by
\citet[hereafter B99]{boesgaard1999} showed that the slope of the Be
vs. O trend in logarithmic scale is about 1.5, which makes the Be
production scenario more complicated and confusing. In addition, the
exact slope of the Be vs. O relationship depends on which oxygen
indicator is used (see discussion in Sect.~\ref{oxygen}).

If Be is produced in a primary process and the cosmic rays were
transported globally across the early Galaxy, Be abundances should
show a very small scatter at a given time. This makes the Be
abundance an ideal cosmic chronometer \citep{suzuki2001}.
\citet{pasquini2004,pasquini2007} found that Be abundances in
globular clusters NGC\,6397 and NGC\,6752 are very similar to that
of the field stars with the same Fe abundances. Furthermore, the
derived ages from Be abundances based on the model of
\citet{valle2002} are in excellent agreement with the ages
determined by main-sequence fitting. They suggested that Be is
produced in primary spallation of cosmic rays acting on a Galactic
scale, and therefore can be used as a cosmochronometer. However, B99
and \citet{boesgaard2006} found strong evidences for intrinsic
spread of Be abundances at a given metallicity, which may indicate
that there is also local enrichment of Be in the Galaxy. But
interestingly, \citet{pasquini2005} found that stars belonging
to the the accretion and dissipative populations (see
\citealt{gratton2003} for the exact definitions for these two
kinematical classes) are neatly separated in the [O/Fe] vs.
$\log$(Be/H) diagram, and especially, the accretion component shows
a large scatter in the [O/Fe] vs. $\log$(Be/H) diagram. They
proposed that most of the scatter in the Be vs. Fe (O) trend may be
attributed to the intrinsic spread of Fe and O abundances (probably
due to the inhomogeneous enrichment in Fe and O of the protogalactic
gas), rather than Be.

In this work, we present Be abundances of 25 metal-poor stars, for
most of which the Be abundances are derived for the first time.
Oxygen abundances are also determined from both \mbox{[O\,{\sc i}]}
forbidden line and \mbox{O\,{\sc i}} triplet lines to investigate
the chemical evolution of Be with O in the Galaxy. In Sect.~\ref{od}
we briefly describe the observations and data reduction. The adopted
model atmosphere and stellar parameters are discussed in
Sect.~\ref{ms}. Sect.~\ref{au} deals with the abundances
determinations and uncertainties. Sect.~\ref{rd} presents the
results and discussions, and the conclusions are given in the last
section.

\section{Observations and data reduction}\label{od}

Our analysis are based on the high resolution and high
signal-to-noise ratio spectra of 25 metal-poor main sequence and
subgiant stars from the archive database of observations obtained
with UVES, the Ultraviolet and Visual Echelle Spectragraph
\citep{dekker2000} at the ESO VLT 8\,m Kueyen telescope. The spectra
were obtained during two observation runs: April 8--12, 2000 and April
10--12, 2001 (Programme ID 65.L-0507 and 67.D-0439), both with standard
Dichroic {\#}1 setting in the blue and red arms. The blue arm spectra
ranged from 3050 to 3850\,{\AA} with a resolution of 48\,000, while
the red arm spectra ranged from 4800 to 6800\,{\AA} with a resolution
of 55\,000.

The spectra were reduced using the standard {\sc eso midas} package.
Reduction procedure includes location of echelle orders, wavelength
calibration, background subtraction, flat-field correction, order
extraction, and continuum normalization.

\section{Model atmospheres and stellar parameters}\label{ms}
\begin{table}
\centering
\caption{Stellar parameters adopted in the analysis.}
\label{parameter}
\begin{tabular}{lccccc}\hline
Star & $T_{\rmn{eff}}$ & $\log g$ & [Fe/H] & $\xi$ & Mass \\
                      & K      & cgs    &    dex    &  km\,s$^{-1}$ & $\mathcal{M}_{\sun}$ \\
\hline
         HD\,76932      &  5890  &  4.12  &  $-0.89$  &  1.2 &  0.91 \\
         HD\,97320      &  6030  &  4.22  &  $-1.20$  &  1.3 &  0.81 \\
         HD\,97916      &  6350  &  4.11  &  $-0.88$  &  1.5 &  1.03 \\
         HD\,103723     &  6005  &  4.23  &  $-0.82$  &  1.3 &  0.87 \\
         HD\,106038     &  5990  &  4.43  &  $-1.30$  &  1.2 &  0.81 \\
         HD\,111980     &  5850  &  3.94  &  $-1.11$  &  1.2 &  0.83 \\
         HD\,113679     &  5740  &  3.94  &  $-0.70$  &  1.2 &  0.94 \\
         HD\,121004     &  5720  &  4.40  &  $-0.73$  &  1.1 &  0.80 \\
         HD\,122196     &  5975  &  3.85  &  $-1.74$  &  1.5 &  0.81 \\
         HD\,126681     &  5595  &  4.53  &  $-1.17$  &  0.7 &  0.71 \\
         HD\,132475     &  5705  &  3.79  &  $-1.50$  &  1.4 &  0.88 \\
         HD\,140283     &  5725  &  3.68  &  $-2.41$  &  1.5 &  0.79 \\
         HD\,160617     &  5940  &  3.80  &  $-1.78$  &  1.5 &  0.86 \\
         HD\,166913     &  6050  &  4.13  &  $-1.55$  &  1.3 &  0.77 \\
         HD\,175179     &  5780  &  4.18  &  $-0.74$  &  1.0 &  0.85 \\
         HD\,188510     &  5480  &  4.42  &  $-1.67$  &  0.8 &  0.55 \\
         HD\,189558     &  5670  &  3.83  &  $-1.15$  &  1.2 &  0.95 \\
         HD\,195633     &  6000  &  3.86  &  $-0.64$  &  1.4 &  1.11 \\
         HD\,205650     &  5815  &  4.52  &  $-1.13$  &  1.0 &  0.77 \\
         HD\,298986     &  6085  &  4.26  &  $-1.33$  &  1.3 &  0.81 \\
CD\,$-$30$\degr$18140   &  6195  &  4.15  &  $-1.87$  &  1.5 &  0.79 \\
CD\,$-$57$\degr$1633    &  5915  &  4.23  &  $-0.91$  &  1.2 &  0.84 \\
         G\,013-009     &  6270  &  3.91  &  $-2.28$  &  1.5 &  0.80 \\
         G\,020-024     &  6190  &  3.90  &  $-1.92$  &  1.5 &  0.83 \\
         G\,183-011     &  6190  &  4.09  &  $-2.08$  &  1.5 &  0.77 \\
\hline
\end{tabular}
\end{table}

We adopted the one dimensional line-blanketed local thermodynamic
equilibrium (LTE) atmospheric model MAFAGS \citep{fuhrmann1997} in
our analysis. This model utilizes the \citet{kurucz1992} ODFs but
rescales the iron abundance by $-0.16$\,dex to match the improved
solar iron abundance of $\log\varepsilon\rmn{(Fe)}=7.51$
\citep{anders1989}. Individual models for each star were computed
with $\alpha$-enhancement of 0.4\,dex if $\rmn{[Fe/H]}<-0.6$ and
with the mixing length parameter $l/H_{\rmn{p}}=0.5$ to get
consistent temperatures from the Balmer lines.

\begin{figure*}
\centering
\includegraphics[width=\textwidth]{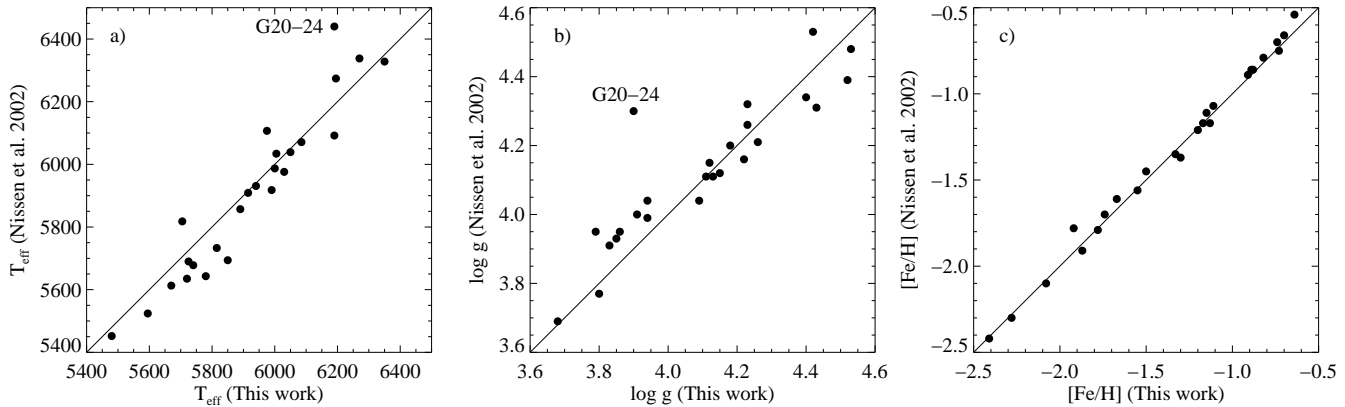}
\caption{Comparison of effective temperature, surface gravity and
iron abundance with \citet{nissen2002}.}
\label{para_comp}
\end{figure*}

The effective temperatures were derived by fitting the wings of
H$\alpha$ and H$\beta$ lines, and then averaged.
\citet{nissen2002} studied oxygen abundances of a large sample
stars, which includes all the objects investigated in this work
(actually our sample is a subset of those employed by
\citealt{nissen2002}). They determined the effective temperatures
from the $b-y$ and $V-K$ colour indices based on the infrared flux
method (IRFM) calibrations of \citet*{alonso1996}. As shown by
Fig.~\ref{para_comp}a, the agreement between the two sets of
temperatures are good for most of the stars with a mean difference
of $15\pm89$\,K. For the star G\,020-024, \citet{nissen2002} gave
a temperature of 6440\,K, which is 250\,K higher than ours.
Recently, \citet{asplund2006} derived a temperature of 6247\,K for
G\,020-024 based on H$\alpha$ profile fitting, which is very close
to ours. \citet{nissen2002} likely overestimated the reddening of
this star \citep{asplund2006}.

The surface gravities were determined from the fundamental relation
\begin{equation}
\log\frac{g}{g_{\sun}}=\log\frac{\mathcal{M}}{\mathcal{M}_{\sun}}
+4\log\frac{T_{\rmn{eff}}}{T_{\rmn{eff},\sun}}+0.4(M_{\rmn{bol}}-M_{\rmn{bol},\sun})
\end{equation}
and
\begin{equation}
M_{\rmn{bol}}=V+BC+5\log\pi+5
\end{equation}
where $\mathcal{M}$ is the stellar mass, $M_{\rmn{bol}}$ is the
absolute bolometric magnitude, $V$ is the visual magnitude, $BC$
is the bolometric correction, and $\pi$ is the parallax.

The absolute visual magnitude were directly derived from the
\emph{Hipparcos} parallax \citep{esa1997} if available with a
relative error smaller than 30\,per cent. For two stars
G\,020-024 and G\,183-011, their uncertainties in parallaxes are
larger than 40\,per cent, so only photometric absolute visual
magnitude can be adopted. For G\,183-011 we followed the
$M_{\rmn{V,phot}}=4.08$ from \citet{nissen2002}. While for
G\,020-024 a large difference exists between the result of
\citet{nissen2002} ($M_{\rmn{V,phot}}=4.33$) and that of
\citet{asplund2006} ($M_{\rmn{V,phot}}=3.72$). We adopted the
latter value because \citet{asplund2006} used the spectroscopic
H$\alpha$ index which provides more precise estimate of
interstellar reddening excess than the photometric H$\beta$ index
employed by \citet{nissen2002}. The bolometric correction was
taken from \citet*{alonso1995} and the stellar mass was estimated
by comparing its position in the
$\log(L/L_{\sun})$-$\log T_{\rmn{eff}}$ diagram with the
evolutionary tracks of \citet*{yi2003}. The final $\log g$ values
are given in Table~\ref{parameter}. Our results are
$0.03\pm0.11$\,dex lower than those of \citet{nissen2002} on
average. Good agreement holds on for majority of the stars except
for G\,020-024, whereas a difference of 0.43\,dex exists. This
is mainly due to the very different absolute visual magnitude
adopted by \citet{nissen2002} and us as discussed above. The
difference in $M_{\rmn{V}}$ (0.61\,mag) alone will introduce a
difference of 0.24\,dex to the surface gravity according to
Equation (1).

\begin{table}
\centering
\caption{Fe {\sc ii} lines used to determine the iron abundances.}
\label{feii}
\begin{tabular}{cccc}\hline
Wavelength & $E_{\scriptsize\textrm{low}}$ & $\log gf$ & $\log C_6$ \\
\AA    & eV                          &           &            \\
\hline
4993.350  &  2.79  &  $-3.73$  &  $-32.18$ \\
5100.664  &  2.79  &  $-4.18$  &  $-31.78$ \\
5197.575  &  3.22  &  $-2.27$  &  $-31.89$ \\
5234.631  &  3.21  &  $-2.21$  &  $-31.89$ \\
5325.560  &  3.21  &  $-3.21$  &  $-32.19$ \\
5425.257  &  3.19  &  $-3.27$  &  $-32.19$ \\
6084.110  &  3.19  &  $-3.84$  &  $-32.19$ \\
6149.250  &  3.87  &  $-2.76$  &  $-32.18$ \\
6247.560  &  3.87  &  $-2.33$  &  $-32.18$ \\
6416.928  &  3.87  &  $-2.67$  &  $-32.18$ \\
6432.680  &  2.88  &  $-3.61$  &  $-32.11$ \\
6456.383  &  3.89  &  $-2.09$  &  $-32.18$ \\
\hline
\end{tabular}
\end{table}

Iron abundances were determined from 12 unblended \mbox{Fe\,{\sc
ii}} lines with spectra synthesis method. We adopted the differential
$\log gf$ values with respect to
$\log\varepsilon\rmn{(Fe)}_{\sun}=7.51$ from \citet*{korn2003} and
the van der Waals damping constants from \citet{anstee1991,anstee1995}.
Our final iron abundances are in excellent agreement with those of
\citet{nissen2002}, who also derived the iron abundances from
\mbox{Fe\,{\sc ii} lines}. The mean difference is only
$-0.02\pm0.05$\,dex. The microturbulent velocities were determined by
requiring that the derived [Fe/H] are independent of equivalent widths.

The typical error for our effective temperature is about $\pm80$\,K.
The uncertainty of parallax contributes most to the error of the
surface gravity. The typical relative error of $\pm$15\,per cent in
parallax corresponds to an error of $\pm$0.13\,dex. In addition,
the estimated error of $\pm$0.05\,$\mathcal{M}_{\sun}$ in stellar
mass translates to an error of $\pm$0.02\,dex, while errors of
$\pm$80\,K in effective temperature and $\pm$0.05\,mag in $BC$ each
leads to an uncertainty of $\pm$0.02\,dex. So the total error of
$\log g$ is about $\pm0.15$\,dex. It is already noted that the iron
abundance is insensitive to the effective temperature. The
uncertainty of [Fe/H] is dominated by the error of surface gravity.
A typical error of $\pm$0.15\,dex in $\log g$ results in an error
of about $\pm$0.07\,dex in [Fe/H]. Combined with the line-to-line
scatter of about $\pm$0.03\,dex, the total error of [Fe/H] is about
$\pm$0.08\,dex. And the error for the microturbulent velocity is
estimated to be about $\pm0.2$\,km\,s$^{-1}$.

\section{Abundances and uncertainties}\label{au}

\subsection{Oxygen}\label{oxygen}

As Be is mainly produced by the spallation of
CNO nuclei, oxygen abundances can be a preferred alternative to iron
in investigating the galactic evolution of Be. It is also important
to know the O content when determining Be abundances, because there
are many OH lines around the \mbox{Be\,{\sc ii}} doublet. Therefore,
the oxygen abundances were firstly investigated here.

There are several indicators for oxygen abundance: the ultraviolet
OH, the \mbox{[O\,{\sc i}]} 6300 and 6363\,{\AA}, the infrared
\mbox{O\,{\sc i}} 7774\,{\AA} triplet and the infrared
vibration-rotation OH lines. However, OH molecules and \mbox{O\,{\sc
i}} atoms in the lower state of the 7774\,{\AA} transitions are
minority species compared to total number of oxygen atoms, thus are
very sensitive to the adopted stellar parameters, such as the
effective temperature and surface gravity. Moreover, line formations
are far from LTE for either the ultraviolet OH
\citep{hinkle1975,asplund2001} or the \mbox{O\,{\sc i}} 7774\,{\AA}
triplet lines \citep{kiselman2001}. In contrast, \mbox{[O\,{\sc i}]}
lines are formed very close to LTE and nearly all the oxygen atoms
in the photosphere of dwarf and giant stars are in the ground
configuration which provide the lower and upper level of the
forbidden lines \citep{nissen2002}. Therefore, it is believed that
the \mbox{[O\,{\sc i}]} line is the most reliable indicator for
oxygen abundances, but the difficulty is that the \mbox{[O\,{\sc i}]}
lines are very weak in dwarf and subgiant stars.

\begin{figure}
\centering
\includegraphics[width=0.4\textwidth]{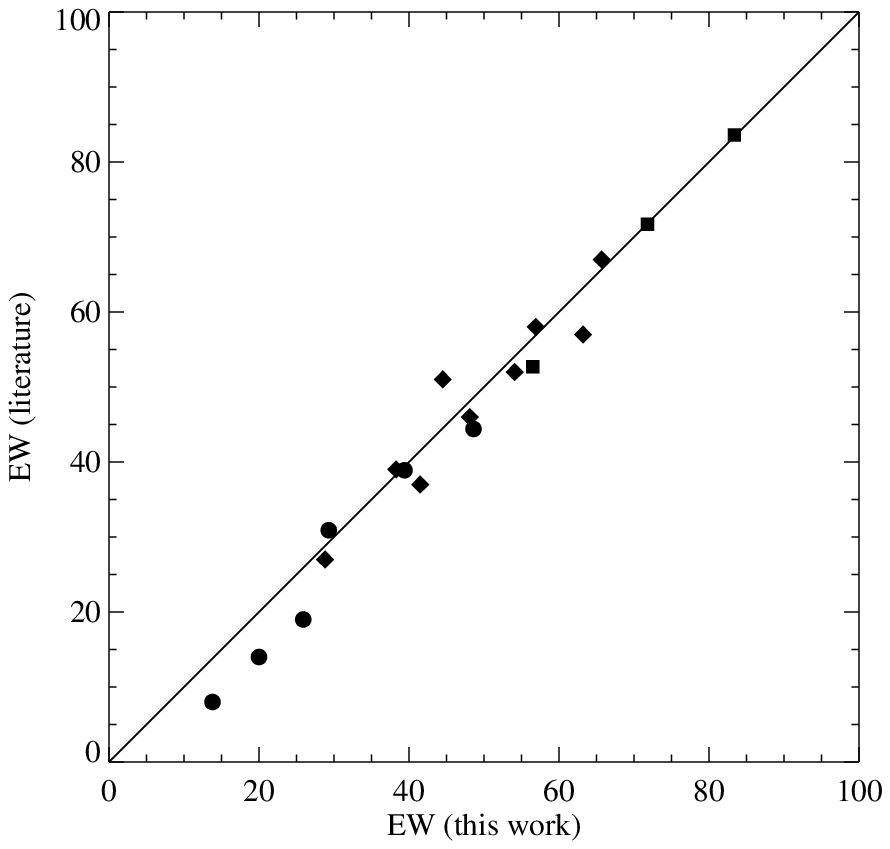}
\caption{Comparison of \mbox{O\,{\sc i}} triplet equivalent widths between
this work and the literatures (\emph{circles}: \citealt{jonsell2005};
\emph{diamonds}: \citealt{nissen1997}; \emph{squares}: \citealt{boesgaard1993}).}
\label{ew_comp}
\end{figure}

High resolution and high sinal-to-noise ratio spectra covering the
infrared \mbox{O\,{\sc i}} triplet for 10 stars were available from
the archived VLT/UVES spectra database. We re-reduced the spectra and
measured the equivalent widths. Since for six of these ten stars
O abundances from the \mbox{O\,{\sc i}} triplet were previously
derived by other authors, we compare our measurements with those
found in literature in Fig.~\ref{ew_comp}. It can be seen that the
agreement is good on the whole. For the rest 15 stars, we collected
their \mbox{O\,{\sc i}} 7774\,{\AA} triplet equivalent widths from
the literatures directly. Oxygen abundances were computed with the
$\log gf=0.369$, 0.223, and 0.002 from \citet*{wiese1996} in LTE
first. Then non-LTE corrections were applied according to the
results of \cite{takeda2003}.

In addition, \citet{nissen2002} measured the equivalent widths of
\mbox{[O\,{\sc i}]} 6300\,{\AA} line for 18 main-sequence and
subgiant stars, of which 15 stars are included in our sample. They
performed the measurement in a very careful fashion, including
removing the possible blending telluric O$_2$ and H$_2$O lines with
the observed spectra of rapidly rotating B-type stars and
subtracting the equivalent width of the blending \mbox{Ni\,{\sc i}}
line at 6300.339\,{\AA}. The typical error for the equivalent width
of \mbox{[O\,{\sc i}]} 6300\,{\AA} line was estimated to be about
only $\pm0.3$\,m{\AA}. For these 15 stars, we also determined their
oxygen abundances with the equivalent widths of \mbox{[O\,{\sc i}]}
6300\,{\AA} line from \citet{nissen2002} using the accurate
oscillator strength $\log gf=-9.72$ from \citet*{allendeprieto2001}.

Finally, we derived the oxygen abundances with \mbox{O\,{\sc i}}
7774\,{\AA} triplet for all the 25 sample stars and with
\mbox{[O\,{\sc i}]} 6300\,{\AA} line for 15 stars, which are given
in Table~\ref{abun} (the reference solar O abundance is
$\log\varepsilon\rmn{(O)}=8.77$ computed from \mbox{[O\,{\sc i}]}
6300\,{\AA} line using the equivalent width of 4.1\,m{\AA} from
\citealt{nissen2002}).

Oxygen abundance from the weak \mbox{[O\,{\sc i}]} 6300\,{\AA} line
is not sensitive to stellar parameters. Its uncertainty is
dominated by the error in equivalent width. Normally, a typical
error of $\pm$0.3\,m{\AA} in equivalent width corresponds on average
to an error of $\pm$0.1\,dex in oxygen abundance. For the infrared
\mbox{O\,{\sc i}} triplet, errors of $\pm$80\,K in effective
temperature and $\pm$0.15\,dex in gravity each translates to an
error of $\pm$0.05\,dex in oxygen abundance. The uncertainties in
iron abundance and microturbulence nearly have no effect on [O/Fe].
A typical error of $\pm$3\,m{\AA} in equivalent width corresponds to
an error of $\pm$0.04\,dex. In total, the error of [O/Fe] from
\mbox{O\,{\sc i}} triplet is around $\pm$0.08\,dex. The uncertainty
stated above is a random one, and the systematic error can be much
higher, as can be seen from the differences of O abundances between
\mbox{[O\,{\sc i}]} forbidden line and \mbox{O\,{\sc i}} triplet
lines discussed below.

\begin{figure}
\centering
\includegraphics[width=0.47\textwidth]{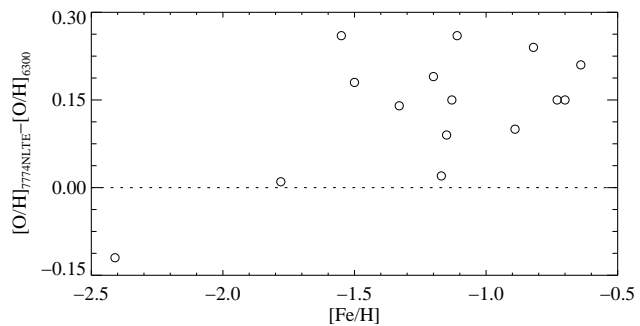}
\caption{\label{6300_7774_comp}Comparison of O abundances from
\mbox{[O\,{\sc i}]} 6300\,{\AA} line with those from \mbox{O\,{\sc
i}} 7774\,{\AA} lines.}
\end{figure}

As there are 15 stars with oxygen abundances from both
\mbox{[O\,{\sc i}]} 6300\,{\AA} and \mbox{O\,{\sc i}} 7774\,{\AA}
lines, it is interesting to investigate whether these two
indicators give consistent oxygen abundances.
Fig.~\ref{6300_7774_comp} shows the differences of O abundances based
on \mbox{[O\,{\sc i}]} 6300\,{\AA} and \mbox{O\,{\sc i}} 7774\,{\AA}
lines. We find that, on average, O abundances from \mbox{O\,{\sc i}}
7774\,{\AA} lines with NLTE corrections are $0.14\pm0.10$\,dex higher
than those from \mbox{[O\,{\sc i}]} 6300\,{\AA} line for the 15 sample
stars, which means that these two indicators are not consistent in our
study. \citet{nissen2002} found that the mean difference between O
abundance from the forbidden and permitted lines is only 0.03\,dex
for five of their stars, and they concluded that these two
indicators produce consistent O abundances. As a test, we reanalyse
the O abundances of \citet{nissen2002}'s sample using their stellar
parameters and equivalent widths. The only differences are the
adopted model atmospheres and the NLTE corrections for the
\mbox{O\,{\sc i}} triplet lines. We found that, for the forbidden
line, our abundances are almost the same as those from
\citet{nissen2002}, but for the triplet lines, our mean LTE O
abundance is about 0.06\,dex larger than that of \citet{nissen2002},
and the NLTE correction for the triplet lines from \citet{takeda2003}
is 0.06\,dex lower than theirs. These two factors lead to a total
difference of 0.12\,dex for the permitted lines. Therefore, the
different model atmospheres (MAFAGS vs. MARCS) and NLTE corrections
are responsible for the differences. Recently,
\citet{garciaperez2006} determined O abundances for 13 subgiant stars
with \mbox{[O\,{\sc i}]}, \mbox{O\,{\sc i}} and OH lines. They
followed exactly the same method as \citet{nissen2002}, but their
results showed that O abundances based on \mbox{O\,{\sc i}} triplet
are on average $0.19\pm0.22$\,dex higher than that from the forbidden
line.

\subsection{Beryllium}

\begin{figure}
\centering
\includegraphics[width=0.47\textwidth]{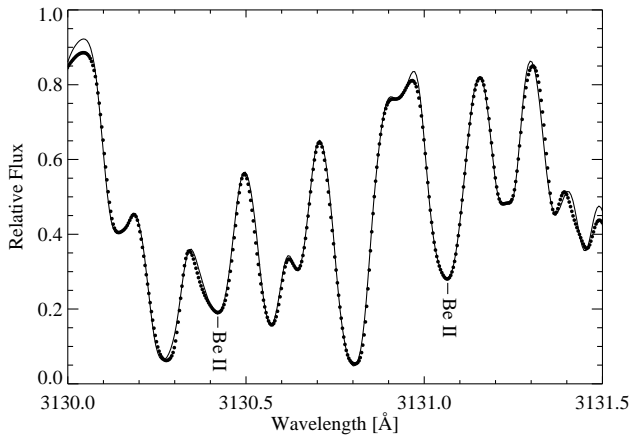}
\caption{Spectral synthesis of the \mbox{Be\,{\sc ii}} doublet region for the KPNO Solar Flux Atlas data.}
\label{be_sun}
\end{figure}

Beryllium abundances were derived by spectra synthesis of the
\mbox{Be\,{\sc ii}} 3130\,{\AA} resonance doublet region using the
{\sc idl/fortran siu} software package of Reetz (1993). It is well
known that this spectral region is rich with atomic and molecular
lines for solar-type stars, which results in substantial line
absorption and a deficit of continuum. We firstly computed the
synthetic solar spectrum around \mbox{Be\,{\sc ii}} doublet region
based on the line list carefully compiled and tested by
\citet{primas1997}, and then compared them with the integrated solar
flux atlas of \citet{kurucz1984}. Some changes were made in order to
make the theoretical solar spectrum match the \citet{kurucz1984}
solar flux atlas best. The major change made to the
\citet{primas1997} line list is that we increased the $\log gf$ of
the \mbox{Mn\,{\sc ii}} 3131.017\,{\AA} line by 1.72\,dex instead of
adding a predicted \mbox{Fe\,{\sc i}} line at 3131.043\,{\AA}.
Similar adjustment was adopted by \citet*{king1997}. Based on this
adjusted line list, we reproduced the solar flux atlas best with
$\rmn{A(Be)}=1.12$, which is in good agreement with the result of
$\rmn{A(Be)}=1.15\pm0.20$ derived by \citet*{chmielewski1975}.
\citet{balachandran1998} found that the standard UV
continuous opacity of the sun need to be multiplied by a factor of
1.6 in order to get consistent oxygen abundances from the UV and IR
OH lines. With the increased UV continuous opacity, they determined
the solar Be abundance to be 1.40, which is very close to the
meteoritic value 1.42. \citet{bell2001} proposed later that the
`missing' UV opacity could be accounted for by the \mbox{Fe\,{\sc
i}} bound-free transitions. However, until now there is no confirmed
evidence about this. But one should keep in mind the `missing' UV
opacity problem. If it does exist, the Be vs. Fe (O) trend of this
work and all the previous work based on the standard UV opacity might
change.

In addition to some strong OH lines, a strong \mbox{Ti\,{\sc ii}}
line at 3130.810\,{\AA} also presents in the \mbox{Be\,{\sc ii}}
region. In order to minimize its effect on the beryllium abundance
as well as to provide a constraint on the location of continuum, we
derived the Ti abundances for our sample stars from the
\mbox{Ti\,{\sc i}} 5866.461, 6258.110, and 6261.106\,{\AA} lines.
The oscillator strengths for these lines are differentially adjusted
to produce the solar Ti abundance $\log\varepsilon\rmn{(Ti)}=4.94$.
For five very metal-poor stars with $\rmn{[Fe/H]}<-1.8$, the
\mbox{Ti\,{\sc i}} lines are too weak to be detected, thus a common
value of $\rmn{[Ti/Fe]}=0.35$ \citep{magain1989} was adopted. As a
matter of fact, due to the metal deficiency of these stars, the line
blending and continuum normalization were much less problematic than
solar-type stars. The abundances of other elements, which are less
critical for Be abundance determination, were adopted by scaling a
solar composition. Beryllium abundances were
then determined by varying the value of Be to best fit the observed
line profiles. It was reported by \citet*{garcialopez1995} and
\citet{kiselman1996} that non-LTE effects for \mbox{Be\,{\sc ii}}
doublet are insignificant, normally smaller than 0.1\,dex. We took
the weaker 3131.066\,{\AA} line as the primary abundance indicator
because it's less blended compared to the stronger 3130.421\,{\AA}
component.

The uncertainties of Be abundances were estimated from the errors of
stellar parameters and pseudo-continuum location. An error of
$\pm$0.15\,dex in surface gravity implies uncertainties of
$\pm$(0.06--0.09)\,dex, while an uncertainty of $\pm$0.08\,dex in
[Fe/H] corresponds to an error of $\pm$0.06\,dex. Errors due to
effective temperature and microturbulence are always within
$\pm$0.04\,dex in total. The error in continuum location was
estimated to be less than five per cent in the worst case, which
results in an error of $\pm$0.15\,dex at most. The final errors for
each star are given in Table~\ref{abun}.
\begin{figure*}
\centering
\includegraphics[width=0.9\textwidth]{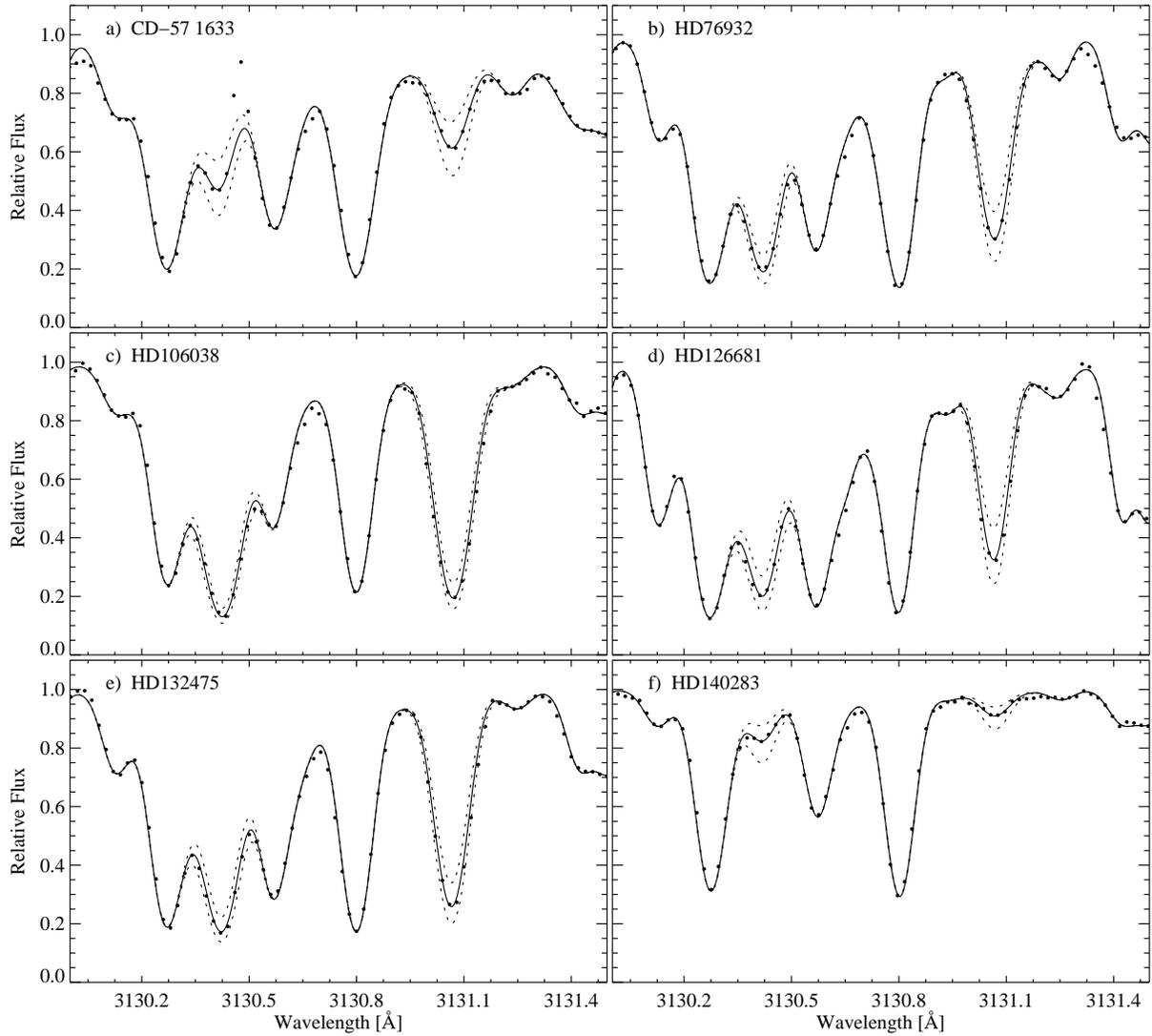}
\caption{\label{be_sample} Spectrum synthesis for six representative
stars. The dots are the observational data, the solid line is the
best-fit synthesis, and the dotted lines are synthetic spectra with
Be abundances of $\pm0.2$\,dex relative to the best fit.}
\end{figure*}

\begin{table*}
\begin{minipage}{150mm}
\caption{Abundances and population membership.}
\label{abun}
\begin{tabular}{lcccccrrcc}\hline
Star & [Fe/H] & \multicolumn{3}{c}{[O/H]} & [Ti/Fe] & A(Li) & A(Be) & $\sigma$(Be) & Pop.$^b$ \\
     &        & 6300 & 7774 LTE$^a$ & 7774 $n$-LTE & & & & & \\
\hline
         HD\,76932        & $-0.89$ & $-0.46$  & $-0.24^{(5)}$ & $-0.36$ & 0.24     & 2.00    & $ 0.73  $   & 0.14      & 0 \\
         HD\,97320        & $-1.20$ & $-0.85$  & $-0.56^{(6)}$ & $-0.66$ & 0.21     & 2.32    & $ 0.43  $   & 0.17      & 0 \\
         HD\,97916        & $-0.88$ & $\cdots$ & $-0.06^{(1)}$ & $-0.27$ & 0.23     & $<1.23$ & $ <-0.76$   & $\cdots$  & 1 \\
         HD\,103723       & $-0.82$ & $-0.68$  & $-0.33^{(6)}$ & $-0.44$ & 0.15     & 2.22    & $ 0.51  $   & 0.17      & 1 \\
         HD\,106038       & $-1.30$ & $\cdots$ & $-0.62^{(3)}$ & $-0.70$ & 0.21     & 2.55    & $ 1.37  $   & 0.12      & 1 \\
         HD\,111980       & $-1.11$ & $-0.76$  & $-0.38^{(6)}$ & $-0.50$ & 0.29     & 2.31    & $ 0.67  $   & 0.13      & 1 \\
         HD\,113679       & $-0.70$ & $-0.43$  & $-0.15^{(6)}$ & $-0.28$ & 0.32     & 2.05    & $ 0.87  $   & 0.12      & 1 \\
         HD\,121004       & $-0.73$ & $-0.42$  & $-0.19^{(2)}$ & $-0.27$ & 0.28     & $<1.18$ & $ 0.94  $   & 0.15      & 1 \\
         HD\,122196       & $-1.74$ & $\cdots$ & $-1.07^{(6)}$ & $-1.17$ & 0.28     & 2.28    & $-0.51  $   & 0.14      & 0 \\
         HD\,126681       & $-1.17$ & $-0.72$  & $-0.65^{(4)}$ & $-0.70$ & 0.30     & 1.48    & $ 0.90  $   & 0.12      & 0 \\
         HD\,132475       & $-1.50$ & $-1.09$  & $-0.82^{(4)}$ & $-0.91$ & 0.27     & 2.23    & $ 0.62  $   & 0.13      & 1 \\
         HD\,140283       & $-2.41$ & $-1.61$  & $-1.65^{(3)}$ & $-1.73$ & $\cdots$ & 2.16    & $-0.94  $   & 0.14      & 1 \\
         HD\,160617       & $-1.78$ & $-1.34$  & $-1.24^{(3)}$ & $-1.33$ & 0.23     & 2.25    & $-0.41  $   & 0.12      & 1 \\
         HD\,166913       & $-1.55$ & $-1.16$  & $-0.81^{(4)}$ & $-0.90$ & 0.30     & 2.32    & $ 0.27  $   & 0.14      & 0 \\
         HD\,175179       & $-0.74$ & $\cdots$ & $-0.13^{(6)}$ & $-0.24$ & 0.32     & $<0.87$ & $ 0.73  $   & 0.15      & 0 \\
         HD\,188510       & $-1.67$ & $\cdots$ & $-0.98^{(5)}$ & $-1.02$ & 0.31     & 1.48    & $-0.25  $   & 0.13      & 0 \\
         HD\,189558       & $-1.15$ & $-0.73$  & $-0.54^{(1)}$ & $-0.64$ & 0.25     & 2.24    & $ 0.64  $   & 0.14      & 0 \\
         HD\,195633       & $-0.64$ & $-0.55$  & $-0.16^{(6)}$ & $-0.34$ & 0.06     & 2.25    & $ 0.53  $   & 0.18      & 2 \\
         HD\,205650       & $-1.13$ & $-0.69$  & $-0.48^{(4)}$ & $-0.54$ & 0.21     & 1.70    & $ 0.51  $   & 0.19      & 0 \\
         HD\,298986       & $-1.33$ & $-0.93$  & $-0.70^{(6)}$ & $-0.79$ & 0.15     & 2.26    & $-0.04  $   & 0.12      & 1 \\
CD\,$-$30$\degr$18140     & $-1.87$ & $\cdots$ & $-1.09^{(3)}$ & $-1.18$ & $\cdots$ & 2.21    & $-0.35  $   & 0.15      & 1 \\
CD\,$-$57$\degr$1633      & $-0.91$ & $\cdots$ & $-0.51^{(6)}$ & $-0.60$ & 0.01     & 2.15    & $ 0.31  $   & 0.18      & 1 \\
         G\,013-009       & $-2.28$ & $\cdots$ & $-1.54^{(3)}$ & $-1.65$ & $\cdots$ & 2.21    & $-0.84  $   & 0.13      & 1 \\
         G\,020-024       & $-1.92$ & $\cdots$ & $-1.19^{(3)}$ & $-1.30$ & $\cdots$ & 2.19    & $-0.72  $   & 0.17      & 1 \\
         G\,183-011       & $-2.08$ & $\cdots$ & $-1.27^{(6)}$ & $-1.36$ & $\cdots$ & 2.21    & $-0.61  $   & 0.14      & 1 \\
\hline
\end{tabular}
$^a$~Sources of equivalent width for O {\sc i} triplet: (1) \citet{cavallo1997}, (2) \citet{nissen1997}, (3) \citet{nissen2002},
(4) \citet{gratton2003}, (5) \citet{jonsell2005}, (6) measured from archival UVES spectra (Programme ID 65.L-0507).

$^b$~Population membership: 0 -- dissipative component; 1 -- accretion component; 2 -- thin disc.
\end{minipage}
\end{table*}

Be abundances for several stars of our sample have been published
by other authors and they are summarized in Table~\ref{be_compare}.
We can see that, though relatively large scatter
exists, Be abundances from different researches are not different
within uncertainties for majority of the stars. The exceptions are
HD\,160617 and HD\,189558. Our Be abundance of HD\,160617 is
0.5\,dex higher than that of \citet{molaro1997}. This difference
is mostly due to the different stellar parameters adopted by
\citet{molaro1997} and us. Their effective temperature and surface
gravity are 276\,K and 0.51\,dex lower than ours, respectively,
which result in a much lower Be abundance. Another exception is
HD\,189558, where a difference of 0.37\,dex in Be abundance exists
between the result of \citet{boesgaard1993} and ours. The slight
differences in the adopted stellar parameters could not produce
such a large difference. \citet{rebolo1988} derived a Be
abundance of $\log N\rmn{(Be/H)}=-12.0\pm0.4$\,dex for this star
with similar stellar parameters. It is 1\,dex lower than the result
of \citet{boesgaard1993}. We noted that \citet{boesgaard1993}
determined Be abundances by measuring the equivalent width of
\mbox{Be\,{\sc ii}} doublet, which is very sensitive to the location
of the continuum. It is probably that they overestimated the
continuum. Moreover, both the spectra of \citet{molaro1997}
and \citet{boesgaard1993} were obtained with 3.6\,m telescopes. The
signal-to-noise ratios around \mbox{Be\,{\sc ii}} region of their
spectra are much lower than that of this work.

\begin{table*}
\begin{minipage}{151mm}
\caption{Comparison of Be abundances with those from literatures.}
\label{be_compare}
\begin{tabular}{lcccccc}\hline
Star            &   (1)  & (2) &  (3)  & (4)  &  (5) &  This work \\
\hline
HD\,76932   &    $-11.04\pm0.11$   &   $-11.45\pm0.18$    &  $-11.21\pm0.21$ &   $-11.17\pm0.05$  &    $\cdots$        &   $-11.27\pm0.14$   \\
HD\,132475  &    $\cdots$          &   $\cdots$           &  $\cdots$        &   $\cdots$         &    $-11.43\pm0.12$ &   $-11.38\pm0.13$   \\
HD\,140283  &    $-12.78\pm0.14$   &   $-13.07\pm0.20$    &  $-12.91\pm0.17$ &   $-13.08\pm0.09$  &    $\cdots$        &   $-12.94\pm0.14$   \\
HD\,160617  &    $\cdots$          &   $\cdots$           &  $-12.90\pm0.27$ &   $\cdots$         &    $\cdots$        &   $-12.41\pm0.12$   \\
HD\,166913  &    $\cdots$          &   $\cdots$           &  $-11.77\pm0.15$ &   $\cdots$         &    $\cdots$        &   $-11.73\pm0.14$   \\
HD\,189558  &    $-10.99\pm0.15$   &   $\cdots$           &  $\cdots$        &   $\cdots$         &    $\cdots$        &   $-11.36\pm0.14$   \\
HD\,195633  &    $-11.21\pm0.07$   &   $\cdots$           &  $\cdots$        &   $\cdots$         &    $-11.34\pm0.11$ &   $-11.47\pm0.18$   \\
\hline
\end{tabular}
References: (1) \citet{boesgaard1993}, (2) \citet{garcialopez1995}, (3) \citet{molaro1997}, (4) B99, (5) \citet{boesgaard2006}.
\end{minipage}
\end{table*}

\subsection{Lithium}

It is well known that beryllium can be destroyed in stars by fusion
reactions at a relatively low temperature (about 3.5 million K). In
order to avoid contamination of our sample from the effect of
depletion process in stars, we also determined the $^7$Li abundances
for our sample stars. Because the destruction temperature for $^7$Li
is lower than that of Be, if $^7$Li is not depleted, Be should not
be depleted either\footnote{For subgiant stars, Li and Be
can be diluted due to the enlargement of the convection zone. In
this case, Li and Be will be diluted by the same percentage.
Nevertheless, Be cannot be depleted more than Li in any case.}.

We adopted the oscillator strengths from the NIST data base, namely
$\log gf=0.002$ and $-0.299$ for the \mbox{Li\,{\sc i}} 6707.76 and
6707.91\,{\AA} lines, respectively. Collisional broadening
parameters describing van der Waals interaction with hydrogen atoms
were taken from \citet*{barklem1998} (see \citealt{shi2007} in
detail). Li abundances were derived by spectra synthesis in LTE.
Results from \citet{asplund2006} and \citet{shi2007} showed that
non-LTE effects for \mbox{Li\,{\sc i}} resonance lines are
insignificant. The typical error for our Li abundance was estimated
to be about 0.1\,dex.

\begin{figure}
\centering
\includegraphics[width=0.47\textwidth]{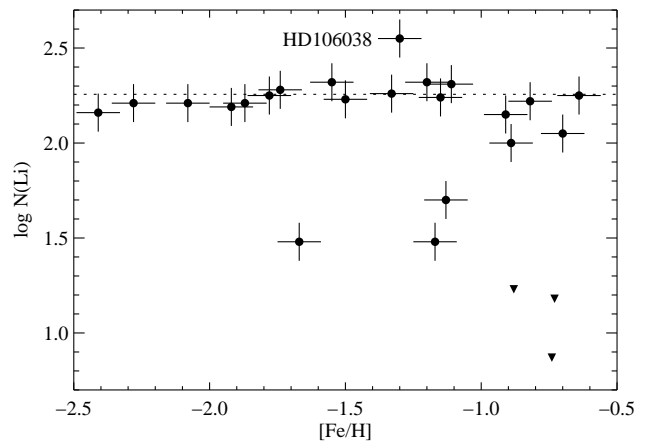}
\caption{\label{li_fe}Li abundances as a function of [Fe/H]. The
filled circles are the determined Li abundances, while the inverse
triangles represent the upper limit.}
\end{figure}

\section{Results and discussion}\label{rd}

\subsection{Be vs. Fe and Be vs. O relation}

\begin{figure}
\centering
\includegraphics[width=0.47\textwidth]{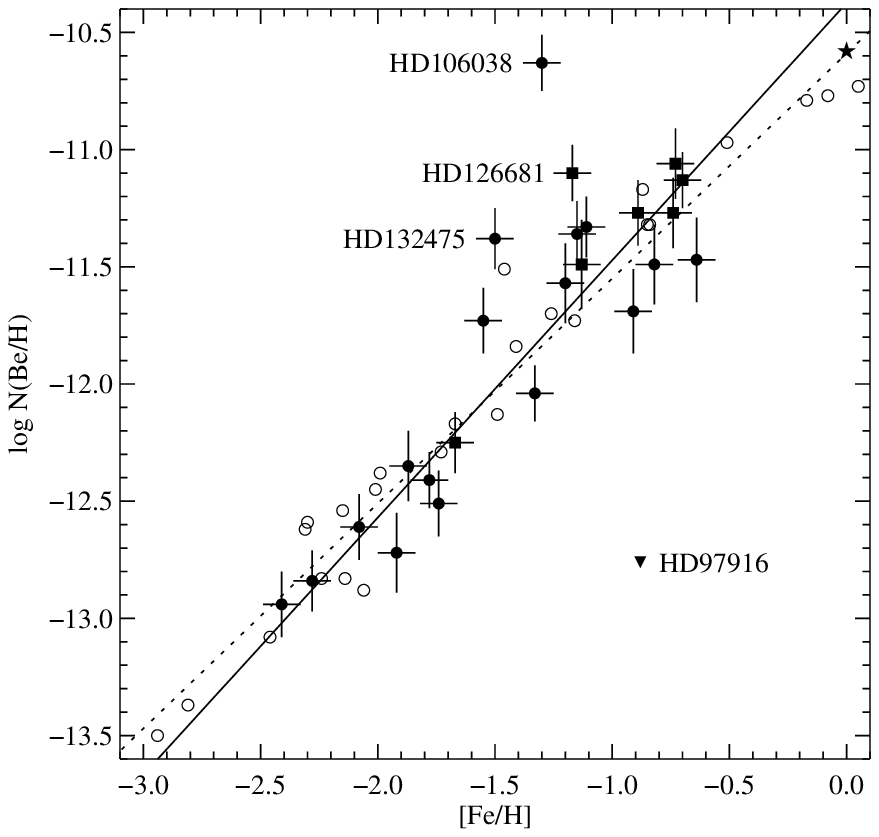}
\caption{\label{be_fe}Be abundances against [Fe/H] (\emph{filled circles}:
stars without $^{7}$Li depletion; \emph{filled squares}: stars with depleted
$^{7}$Li; \emph{filled inverse triangle}: upper limit Be abundance for
HD\,97916; \emph{filled star}: solar meteoritic Be abundance; \emph{open
circles}: data from B99). The solid line is the best
linear fitting (not including HD\,106038, HD\,97916, HD\,132475 and
HD\,126681) with a slope of 1.1 for our Be vs. Fe trend, while the dotted
lines represents the best fitting for the B99 data.}
\end{figure}

\begin{figure*}
\centering
\includegraphics[width=0.9\textwidth]{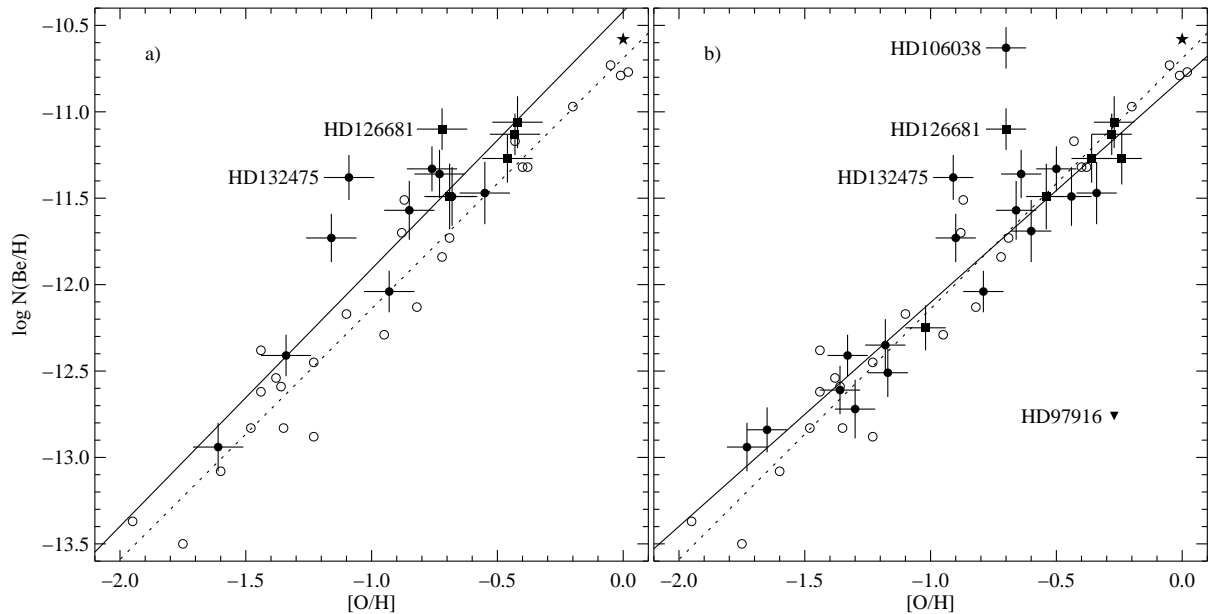}
\caption{\label{be_o}Be abundances as a function of [O/H]. a)
Results from \mbox{[O\,{\sc i}]} forbidden lines. b) Results from
\mbox{O\,{\sc i}} permitted lines. The symbols and lines have the same
meaning as Fig.~\ref{be_fe}. Remember that O abundances of
B99, for both of the two panels, are the mean
abundances based on UV OH and infrared \mbox{O\,{\sc i}} triplet lines.}
\end{figure*}

Encouraged by the agreement of Be abundances between our results and
literatures, we now turn to investigate the chemical evolution of
beryllium in the Galaxy. As we have mentioned before, it is first
necessary to investigate whether some of our stars are depleted in Be.
Fig.~\ref{li_fe} displays the $^7$Li abundances as a function of
metallicity. We can see that six stars in our sample are obviously
depleted in $^7$Li and another two seem to be slightly depleted,
while the star HD\,106038 has an exceptionally high $^7$Li abundance,
about 0.3\,dex higher than the Spite plateau (it also has an
abnormally high Be abundance, see discussion in Sect.~\ref{spec} for
this star). Among those stars with depleted $^7$Li, only one
(HD\,97916, denoted by filled inverse triangle in Fig.~\ref{be_fe}
and Fig.~\ref{be_o}) is also depleted seriously in Be, while the rest
(denoted by filled squares in Fig.~\ref{be_fe} and Fig.~\ref{be_o})
seem to have normal Be abundances at their metallicities.

Excluding HD\,106038 and HD\,97916 (abnormally high Be abundance and
seriously depleted in Be, respectively; these two stars will not be
included in the analysis of Be vs. O trend either), the relation
between Be and Fe abundances can be well represented by linear fitting
\[
\log N\rmn{(Be/H)}=(1.15\pm0.07)\,\rmn{[Fe/H]}-(10.24\pm0.10)
\]
One may note that HD\,132475 and HD\,126681 seem to deviate from the
trend far beyond the uncertainties. If we exclude these two stars
(see discussion in Sect.~\ref{spec}), the relationship will be
\[
\log N\rmn{(Be/H)}=(1.10\pm0.07)\,\rmn{[Fe/H]}-(10.37\pm0.10)
\]
Our result is in reasonable agreement with the result
$\log N\rmn{(Be/H)}=(0.96\pm0.04)\,\rmn{[Fe/H]}-(10.59\pm0.03)$
of B99, considering the relatively smaller
metallicity range of our sample. The overall increase of Be with Fe
suggests that Be is enriched globally in the Galaxy.

As the yields of Be is believed to be correlated with CNO nuclei
directly, it's more meaningful to investigate the relationship
between Be and O abundances. Due to the inconsistent oxygen
abundances based on the forbidden and triplet lines in our
study, it is necessary to investigate their relations with Be
abundance separately. Fig.~\ref{be_o} shows the trend of Be with
O abundances. Again, the Be abundances increase linearly with
increasing [O/H] both for the forbidden line (though with relatively
large scatter partly due to the small sample number) and triplet
lines. The relationships are best represented by
\begin{eqnarray*}
\log N\rmn{(Be/H)}=(1.55\pm0.17)\,\rmn{[O/H]}-(10.29\pm0.15) & [\rmn{O}\,\rmn{\scriptstyle{I}}] \\
\log N\rmn{(Be/H)}=(1.36\pm0.09)\,\rmn{[O/H]}-(10.69\pm0.08) & \rmn{O}\,\rmn{\scriptstyle{I}}
\end{eqnarray*}
If we exclude HD\,132475 and HD\,126681, the relationships will be
\begin{eqnarray*}
\log N\rmn{(Be/H)}=(1.49\pm0.16)\,\rmn{[O/H]}-(10.42\pm0.15) & [\rmn{O}\,\rmn{\scriptstyle{I}}] \\
\log N\rmn{(Be/H)}=(1.30\pm0.08)\,\rmn{[O/H]}-(10.81\pm0.08) & \rmn{O}\,\rmn{\scriptstyle{I}}
\end{eqnarray*}

Our result based on \mbox{O\,{\sc i}} triplet lines is slightly
flatter than the result
$\log N\rmn{(Be/H)}=(1.45\pm0.04)\,\rmn{[O/H]}-(10.69\pm0.04)$
of B99. This can be partly due to our smaller
[Fe/H] range as mentioned above. Besides, the O abundances of
B99 were averaged from the UV OH and infrared
\mbox{O\,{\sc i}} triplet lines. They thought that such a result
(a slope of roughly 1.5 for Be vs. O) is neither consistent with
the secondary process, nor the primary process. However, the
secondary process added with some chemical evolution effects, such
as an outflow of mass from the halo, indicates that there would be
a quadratic relation only at the very lowest metallicities and a
progressive shallowing of the slope to disc metallicities, for
example a slope of 1.5 between $\rmn{[Fe/H]}=-2$ and $-1$ (B99).
So they suggested that this process is most consistent with their
results.

As discussed above, \mbox{[O\,{\sc i}]} forbidden line is the most
reliable O abundance indicator for its insensitivity to the
adopted stellar parameters as well as the non-LTE effects. Our
oxygen abundances from \mbox{[O\,{\sc i}]} 6300\,{\AA} line produces
a `moderate' slope 1.49 for the Be vs. O trend. However,
\citet{nissen2002} studied the effects of 3D model atmospheres on
the derived O abundance from \mbox{[O\,{\sc i}]} forbidden lines.
They found that O abundances based on \mbox{[O\,{\sc i}]} will
decrease if 3D models are applied. Especially, the decreasing
amplitude increases with decreasing metallicity (see fig.~6 and
table~6 of \citealt{nissen2002}). While 3D effects on \mbox{Be\,{\sc
ii}} doublet are negligible according to \citet{primas2000}. This
means that the slope will be closer to one than our present result
when the 3D effects were considered for the \mbox{[O\,{\sc i}]}
forbidden lines. This implies that the Be production scenario is
probably a primary process.

\subsection{Special stars: hints on the Be production scenario}\label{spec}

HD\,106038 is very special for its exceptionally overabundance in Li
and Be as mentioned before. Its Li abundance is 0.3\,dex higher than
the Li plateau, while its Be abundance is about 1.2\,dex higher than
that of normal stars with the same metallicity. The Be abundance of
this star is even similar to the solar meteoritic value. Such a Be-rich
star is extremely rare. \citet{asplund2006} derived $\rmn{A(Li)}=2.49$
and \citet{smiljanic2008} reported a Be abundance of
$\log N\rmn{(Be/H)}=-10.60$ for this star, both in good agreement
with our results. In addition to Li and Be, \citet{nissen1997} showed
that this star also has obviously enhanced abundances of Si, Ni, Y,
and Ba. Based on its special abundance pattern, \citet{smiljanic2008}
suggested that HD\,106038 is most probably formed in the vicinity of
a hypernova (HNe).

In addition to HD\,106038, another two stars, namely HD\,132475 and
HD\,126681, seem to stand out of the Be vs. Fe and Be vs. O trends
distinctly. Their Be abundances are about 0.6 and 0.5\,dex higher,
respectively, than that of most stars with the same metallicities.
\citet{boesgaard2006} also found an abnormally high Be abundance
for HD\,132475 (0.5\,dex above the normal stars at its metallicity).
Their sample included another star BD$+$23$\degr$3912, which has
very similar atmospheric parameters as HD\,132475 but very different
Be abundance. In fact, BD$+$23$\degr$3912 has a Be abundance
matching perfectly the linear Be vs. Fe relation. Combined with
another star HD\,94028 also with excess Be abundance found by B99,
\citet{boesgaard2006} concluded that dispersion in Be abundances
does exist at a given metallicity, which implies a local enrichment
of Be in the Galaxy.

However, \citet{pasquini2005} proposed that such dispersion
could be mostly attributed to the scatter of Fe and O abundances,
rather than Be, as we have mentioned in Sect.~\ref{intro}. Following
\citet{pasquini2005}, we calculated the space velocities using the
method presented by \citet{johnson1987}, and determined the orbital
parameters based on the Galactic mass model of \citet{allen1991} for
our sample stars. Input parameters, such as radial velocities,
parallaxes and proper motions were obtained from the SIMBAD
database. According to the criteria of \citet{gratton2003}, fifteen
stars in our sample belong to the accretion component, nine stars
belong to the dissipative component and one star belongs to the thin
disc.

\begin{figure*}
\centering
\includegraphics[width=0.9\textwidth]{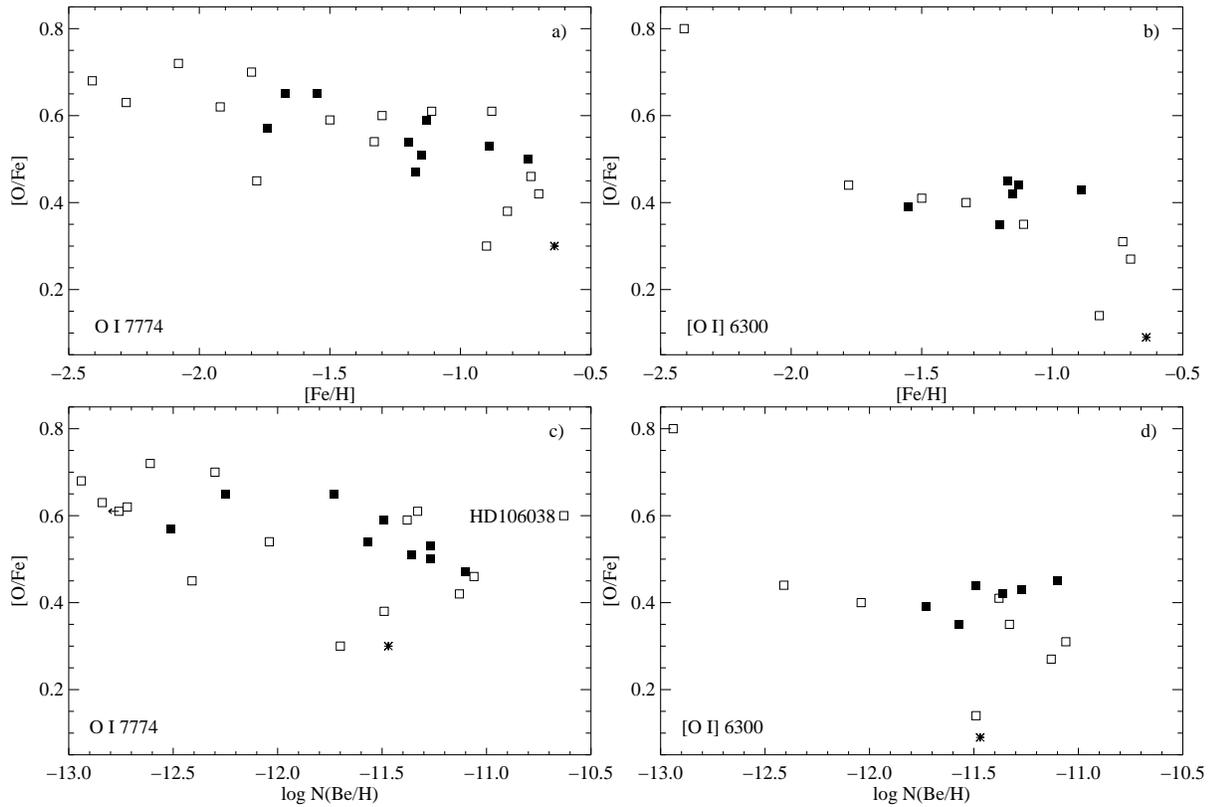}
\caption{\label{kin}a) [O/Fe] vs. [Fe/H] based on \mbox{O\,{\sc i}}
triplet lines. b) [O/Fe] vs. [Fe/H] based on \mbox{[O\,{\sc i}]}
forbidden line. c) [O/Fe] vs. $\log N\rmn{(Be/H)}$ based on
\mbox{O\,{\sc i}} triplet lines. d) [O/Fe] vs. $\log N\rmn{(Be/H)}$
based on \mbox{[O\,{\sc i}]} forbidden line. \emph{Open squares}
represent the accretion component, \emph{filled squares} represent
the dissipative component, and the \emph{asterisk} represents the
only thin disc star HD\,195633 in our sample. The open square with
a \emph{left pointing arrow} represents the upper limit Be abundance
for HD\,97916.}
\end{figure*}

Fig.~\ref{kin} shows [O/Fe] vs. [Fe/H] and [O/Fe] vs. $\log
N\rmn{(Be/H)}$ for our sample stars, based on the results from both
\mbox{[O\,{\sc i}]} forbidden line and \mbox{O\,{\sc i}} triplet
lines. One may find that no clear separation exists between the two
components in the [O/Fe] vs. [Fe/H] diagram, though the accretion
component shows a relatively larger scatter than the dissipative
component. However, in the [O/Fe] vs. $\log N\rmn{(Be/H)}$ diagram,
the two populations are distinctly different, and especially, the
accretion component shows a much larger scatter compared to the
dissipative component. Our results agree well with the findings of
\citet{pasquini2005}. They proposed that such results support the
idea that the formation of the two components took place under
significantly different conditions: an inhomogeneous, rapidly
evolving `halo phase' for the accretion component, and a more
chemically homogeneous, slowly evolving `thick disc phase' for the
dissipative component. The large scatter of the accretion component
in the [O/Fe] vs. $\log N\rmn{(Be/H)}$ diagram may reflect the
inhomogeneous enrichment in oxygen and iron of the halo gas. We note
that, for our Be-rich stars, HD\,106038 and HD\,132475 belong to the
accretion component, while HD\,126681 belongs to the dissipative
component. Another Be-rich star HD\,94028, first discovered by B99
and later confirmed by \citet{boesgaard2006}, is also classified as
a dissipative component star according to the definitions of
\citet{gratton2003}. While the deviation of the accretion component
stars could be due to the inhomogeneous enrichment in Fe and O of
the halo gas, HD\,126681 and HD\,94028 cannot be interpreted in this
way. However, one should keep in mind that stellar kinematics is
only of statistical meaning in describing the Galactic populations.
It has been shown by many previous work that substantial overlap
exists between the halo and thick disc stars. Therefore, it is
dangerous to attribute an individual star to one stellar population
and accordingly derive a firm conclusion. Moreover, we didn't find
any distinct differences in the abundance pattern between
HD\,126681/HD\,94028 and typical halo stars. So the possibility that
dispersion in Be vs. Fe and Be vs. O trend originates from the
inhomogeneous enrichment in Fe and O of the protogalactic gas cannot
be excluded.

As an alternative, the scatter in Be can be interpreted by
the so-called superbubble (SB) model (\citealt{higdon1998,
parizot1999,parizot2000b,ramaty2000}). The basic idea of the SB
model is that repeated SNe occurring in an OB association can
generate a superbubble, in which the CNO nuclei (ejected by SNe)
mixed with some ambient, metal-poor material are accelerated onto
the metal deficient material in the supershell and at the surface of
the adjacent molecular cloud, and then broken into smaller atoms
like Li, Be and B. The produced light elements are then mixed with
other SNe ejecta as well as the ambient, metal-poor gas. However,
such a mixing cannot be perfect, and new stars can form before all
the massive stars explode or the induced light elements production
occurs. As a result, scatter in the abundances of light elements for
a given SB may occur \citep{parizot2000a}. \citet{boesgaard2006}
noted that Na, Mg, Si, Y, Zr, and Ba abundances of HD\,132475 are
typically 0.2\,dex above the mean values of the other stars at that
metallicity according to the results of
\citet{fulbright2000,fulbright2002}. We also noted that Y and Ba
abundances of HD\,126681 are roughly 0.2 and 0.15\,dex higher,
respectively, than the mean values of the remaining sample as found
by \citet{nissen1997}, and the $\alpha$-elements of this star are in
the upper range of their sample. Na and $\alpha$-elements are
typical ejecta of SNe\,II, and Y, Zr, and Ba, though not very
efficiently, can also be produced by the $r$-process in SNe\,II. It
is probably that stars like HD\,132475 and HD\,126681 were formed
from the material that underwent enrichment of light elements and
SNe\,II ejecta but not plenty of dilution process in SBs. In
addition, the SB model also predicts a primary process for the Be
production, which is consistent with our Be vs. O trend.

Except for the possibilities stated above, some other
scenarios for the overabundance of Be can be excluded. It is
unlikely that the Be-rich stars were accreted from the satellite
systems of our Galaxy. \citet*{shetrone2001} and \citet{tolstoy2003}
found that abundance patterns among the dwarf spheroidal (dSph)
galaxies stars are remarkably uniform. We note that
$\alpha$-elements and Y abundances of the dSph stars are obviously
lower than the Be-rich stars. The possibility that the overabundance
of Be in our Be-rich stars could be due to the accretion of a planet
or planetesimals debris can also be excluded. If the excess Be in
our Be-rich stars were accreted from some material having similar
composition as chondrites meteorites, then the mass of the accreted
iron would be even larger than the total mass of iron in the surface
convective zone of the star, which is certainly impossible.

\section{Conclusions}\label{con}

We have derived Be abundances for 25 main sequence and subgiant stars
spanning the range $-2.5$ $<$ [Fe/H] $<$ $-0.5$. Relations between Be
and Fe as well as Be and O are investigated. The Be vs. Fe trend can
be well represented by a linear relation with a slope of 1.1. This
result is in good agreement with that of B99, and
suggests that Be is enriched globally in the Galaxy, as proposed by
\citet{suzuki2001}. Our Be abundances increase linearly with
increasing [O/H] based on both the \mbox{[O\,{\sc i}]} 6300\,{\AA}
and \mbox{O\,{\sc i}} triplet lines, but with slightly different
slopes. O abundances based on \mbox{O\,{\sc i}} triplet gives a slope
of 1.30 between [Be/H] and [O/H]. This is a little flatter than the
result of B99, which may be partly due to different
metallicity range. The most reliable O abundance indicator,
\mbox{[O\,{\sc i}]} forbidden line gives a steeper relationship (a
slope of 1.49). However, this slope will decrease if 3D effects are
took into account according to the results of \citet{nissen2002},
which means that the production process of Be is probably a primary
process.

Moreover, we find some strong evidences for the intrinsic dispersion
of Be abundances at a given metallicity. The special abundance
pattern of HD\,106038, especially its exceptionally high Be
abundance, can be interpreted most consistently only if the material
which formed HD\,106038 was contaminated by the nucleosynthetic
products of a HNe \citep{smiljanic2008}. The deviations of
HD\,132475 and HD\,126681 from the general Be vs. Fe and Be vs. O
trend can be interpreted by the SB model. However, the possibility
that such dispersion originates from the inhomogeneous enrichment in
Fe and O of the protogalactic gas cannot be excluded.

\section*{Acknowledgments}

Thanks goes to the referee Luca Pasquini for constructive suggestions
and comments. This work is supported by the National Natural Science
Foundation of China under grants Nos. 10433010, 10521001 and 10778626.
It has made use of the SIMBAD database, operated at CDS, Strasbourg,
France.

\bsp

\label{lastpage}

\end{document}